\documentclass[pra,twocolumn]{revtex4} 
\usepackage{graphicx}
\usepackage{bm}
\usepackage{amsmath, amssymb}

\usepackage{color}

\begin{document}

\title{Low temperature properties of the infinite-dimensional 
attractive Hubbard model}
\author{Akihisa Koga}
\affiliation{Department of Physics, Tokyo Institute of Technology, 
Tokyo 152-8551, Japan}
\author{Philipp Werner}
\affiliation{Theoretische Physik, ETH Zurich, 8093 Z\"urich, Switzerland}
\date{\today}%

\begin{abstract}
We investigate the attractive Hubbard model in infinite spatial dimensions 
by combining dynamical mean-field theory with 
a strong-coupling continuous-time quantum Monte Carlo method.
By calculating the superfluid order parameter and the density of states, 
we discuss the stability of the superfluid state.
In the intermediate coupling region above the critical temperature,
the density of states exhibits a 
heavy fermion behavior with a quasi-particle peak 
in the dense system, while a dip structure appears in the dilute system.
The formation of the superfluid gap is also addressed.
\end{abstract}
\maketitle

\section{Introduction}
Ultracold atomic systems have attracted a lot of interest
following the demonstration of Bose-Einstein condensation (BEC) 
in a Rb atom system~\cite{Rb}.
Among the many interesting issues, the nature of the superfluid state 
in fermionic systems \cite{Regal}
is a widely studied topic, since a crossover between a weak-coupling 
BCS-type superfluid 
state and a strong-coupling BEC-type superfluid state (BCS-BEC crossover)
has been observed in experiments~\cite{BCSBEC1,BCSBEC2,BCSBEC3}.
Recently, the appearance of a pseudogap in the density of states 
has also been suggested~\cite{JILA},
which stimulates further theoretical investigations on the superfluid state
and related phenomena.

Fermi gas systems have been studied theoretically in much detail and 
it has been clarified that a pseudogap phenomenon indeed appears 
in the BCS-BEC crossover region above 
the critical temperature~\cite{Chen,Tsuchiya,Chien,Su}.
More recently, the gap structure in the superfluid state 
has been discussed~\cite{Pieri,Watanabe}.
It has been pointed out that upon decreasing temperature, 
the pseudogap disappears at a certain temperature 
inside the superfluid phase
and the superfluid gap opens instead~\cite{Watanabe}.
On the other hand, in fermionic optical lattice systems
described by the attractive Hubbard model, 
such dynamical properties at finite temperatures have not been discussed.
It is also important to clarify how the particle density as well as
the interaction strength affect the low energy properties
in the pseudogap region. 
Therefore, it is useful to study the stability of the superfluid state
systematically and to clarify how the gap structure appears 
in the density of states at low temperatures.

In this paper,
we consider the infinite-dimensional attractive Hubbard model
to discuss its low temperature properties. To take 
correlation effects precisely into account,
we combine dynamical mean-field theory 
(DMFT)~\cite{Metzner,Muller,Georges,Pruschke} 
with a continuous-time quantum Monte Carlo (CTQMC) 
method~\cite{solver_review}.
This unbiased technique enables us to discuss 
the stability of the superfluid state quantitatively.
By tuning the particle density, interaction strength, and temperature,
we determine phase diagrams of the system and 
discuss the formation of the gap in the density of states.

The paper is organized as follows.
In \S2, we introduce the attractive Hubbard model and 
briefly summarize our theoretical approach.
We discuss how the superfluid state is realized at low temperatures in \S3.
In \S4, we focus on the dilute system 
in the BCS-BEC crossover region.  
We clarify how the gap structure appears in the density of states 
and discuss the difference of low energy properties 
in lattice and Fermi gas systems.
A brief summary is given in the last section.

\section{Model and Method}
We consider two-component fermions in an optical lattice,
which may be described by the following Hubbard Hamiltonian,
\begin{equation}
\hat{\cal{H}}=-t\sum_{(i,j),\sigma}
c^{\dagger}_{i\sigma}c_{j\sigma}
-U\sum_{i}n_{i\uparrow}n_{i\downarrow},
\label{eq1}
\end{equation}
where $c_{i\sigma}$ ($c^{\dagger}_{i\sigma}$)
is an annihilation (creation) operator of a fermion on the $i$th site
with spin $\sigma (=\uparrow, \downarrow )$, and
$n_{i\sigma}= c^{\dagger}_{i\sigma}c_{i\sigma}$.
$U$ is the onsite attractive interaction and 
$t$ is the transfer integral between sites.
The low-energy properties of the attractive Hubbard model have been studied 
in one dimension~\cite{Lieb,Shiba,Machida,Xianlong,Pour,Fujihara},
two dimensions~\cite{TD1,TD2,Koga}
and infinite
dimensions~\cite{Garg,Toschi,Suzuki,Keller,Capone,Bauer,Freericks,Privitera,KogaQMC1}.
It is known that the superfluid ground state is always realized
in two and higher dimensions, where the BCS-BEC crossover 
has been studied in detail.
However, dynamical properties have not been studied yet in detail 
in the intermediate correlation and temperature region. 
In particular, the question whether and how 
a pseudogap appears in the density of states
above the critical temperature, has not been answered.
In this paper, we systematically investigate the low temperature properties in 
the attractive Hubbard model 
by varying the particle density, interaction strength, 
and temperature.
We then clarify how the gap structure appears in the density of states.

For this purpose, we make use of DMFT~\cite{Metzner,Muller,Georges,Pruschke}.
In DMFT, the original lattice model is mapped to an effective impurity model,
where local particle correlations are accurately taken into account.
The lattice Green's function is obtained via a self-consistency 
condition imposed on the impurity problem.
This treatment is exact in infinite dimensions, and
the DMFT method has successfully been applied to 
strongly correlated fermion systems.
In DMFT, we take into account dynamical correlations
through the frequency-dependent self-energy.
This allows us to discuss the stability of the $s$-wave superfluid state 
more quantitatively than in the static BCS mean-field theory.

The lattice Green's function is given by 
\begin{eqnarray}
\hat{G}^{-1}(k, i\omega_n)=i\omega_n\hat{\sigma}_0+
\left(\mu-\epsilon_k\right)\hat{\sigma}_z-\hat{\Sigma}\left(i\omega_n\right),
\end{eqnarray}
where $\mu$ is the chemical potential,
$\hat{\sigma}_0$ and $\hat{\sigma}_z$ are the identity matrix and
the $z$-component of the Pauli matrix,
$\epsilon_k$ is the dispersion relation for the non-interacting
system, and $\omega_n$ is the Matsubara frequency.
$\hat{\Sigma}(i\omega_n)$ is the local self-energy in the Nambu formalism.
The local lattice Green's function is obtained as 
\begin{eqnarray}
\hat{G}(i\omega_n) = \int dk \hat{G}(k, i\omega_n).
\end{eqnarray}
In this paper, we use a semi-circular density of states,
$\rho(x) = 2/\pi D \sqrt{1-(x/D)^2}$,
where $D$ is the half bandwidth, which corresponds to 
an infinite coordination Bethe lattice.
The self-consistency equation \cite{GeorgesZ} is then given by
\begin{equation}
\hat{G}_{0,\text{imp}}^{-1}(i\omega_n) = i\omega_n \hat{\sigma}_0
+\mu\hat{\sigma}_z-\left(\frac{D}{4}\right)^2\hat{\sigma}_z\hat{G}(i\omega_n)
\hat{\sigma}_z\label{eq:self}.
\end{equation}

There are various numerical methods to solve the effective impurity problem.
To study the attractive Hubbard model systematically,
an unbiased and accurate numerical solver is necessary, 
such as the exact diagonalization\cite{Caffarel,Toschi,Privitera} or 
the numerical renormalization group~\cite{NRG,NRG_RMP,OSakai,Bauer}.
A particularly powerful method for exploring finite temperature properties 
is CTQMC~\cite{solver_review}. 
In our previous paper~\cite{KogaQMC1}, 
we have used the CTQMC method in the continuous-time
auxiliary field formulation~\cite{CTAUX} to study 
the imbalanced attractive Hubbard model.
However, this is a weak-coupling approach, 
which is not suitable for a systematic investigation of 
the low-temperature properties in the strong coupling regime. 
Hence, our previous discussion was 
restricted to the weak coupling region. Here, 
we employ the complementary strong coupling version of 
the CTQMC method~\cite{CTQMC},
which is more efficient in the large $|U|$ region.
We use this method to investigate the attractive Hubbard model both in the 
weak and strong coupling regimes.
Some details of the implementation are explained in the appendix.

In this paper, we use the half bandwidth $D$ as the unit of energy. 
We then calculate the pair potential $\Delta$, double occupancy $d$, 
internal energy $E$, and specific heat $C$, 
which are given~\cite{Toschi} by 
\begin{eqnarray}
\Delta&=&\langle c_{i\uparrow} c_{i\downarrow}\rangle = 
\hat{G}_{12}(\tau=0_+),\\
d&=&\langle n_{i\uparrow} n_{i\downarrow}\rangle,\\
E&=&\left(\frac{D}{2}\right)^2 \int_0^\beta d\tau {\rm Tr}
\left[ \hat{G}(\tau)\hat{\sigma}_z\hat{G}(-\tau)\hat{\sigma}_z\right]+Ud,\\
C&=&\frac{dE}{dT}.
\end{eqnarray}
In addition to these static quantities, we deduce the density of states by
applying the maximum entropy method~\cite{MEM1,MEM2,MEM3}
to the Green's function.
We then discuss how the gap structure appears in the system.

\section{Stability of the superfluid state}

\begin{figure}[b]
\begin{center}
\includegraphics[width=7cm]{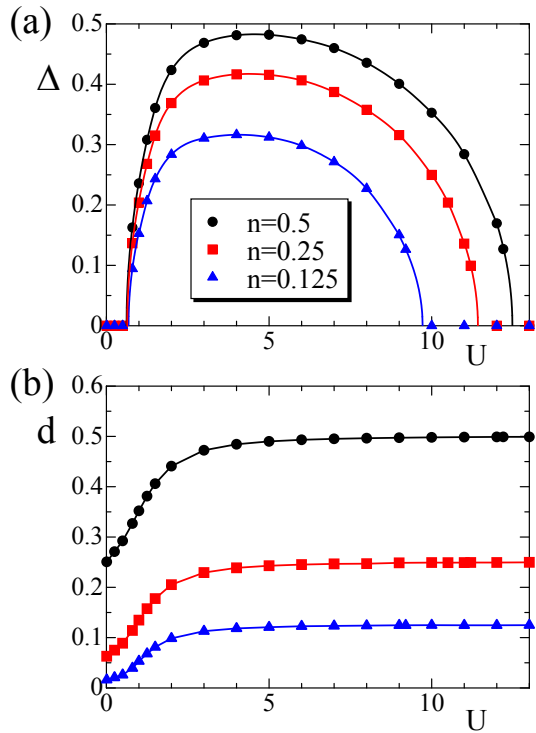}
\caption{Pair potential (a) and double occupancy (b) as a function of 
the interaction at temperature $T=0.02$. 
Circles, squares, and triangles show the results for 
the system with $n=0.5$, $0.25$ and $0.125$, respectively.
}
\label{fig:50}
\end{center}
\end{figure}

We first consider the attractive Hubbard model with different band fillings
to discuss how the superfluid state is realized
at low temperatures~\cite{Toschi,Privitera,Bauer,Garg}.
Figure~\ref{fig:50} shows the pair potential $\Delta$ and 
the double occupancy $d$
at a fixed temperature $T=0.02$. 
In the noninteracting case $(U=0)$, a normal metallic state is realized,
with $\Delta=0$ and $d=n^2$.
Attractive interactions lead to the formation of Cooper pairs 
and an increase in the double occupancy, as shown in Fig.~\ref{fig:50}~(b). 
More precisely, at a certain critical interaction $U_{c1}$, 
a phase transition occurs
to a superfluid state and the pair potential is induced,
as shown in Fig.~\ref{fig:50}~(a).
A cusp singularity appears in the curve of the double occupancy 
although it is not visible on this scale.

\begin{figure}[htb]
\begin{center}
\includegraphics[width=7cm]{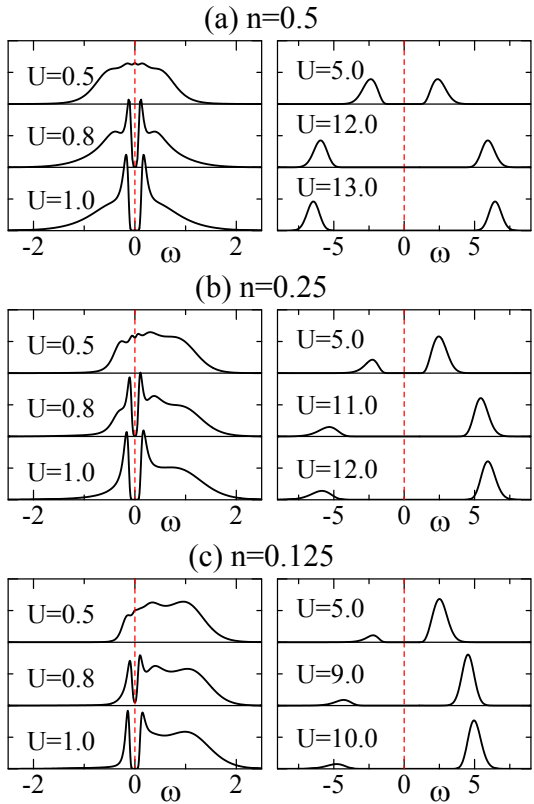}
\caption{Density of states for the systems with $n=0.5$ (a), $0.25$ (b) 
and $0.125$ (c) at $T=0.02$.
}
\label{fig:dos-50}
\end{center}
\end{figure}

To clarify how the phase transition affects dynamical properties,
we show the density of states deduced by the MEM in Fig.~\ref{fig:dos-50}.
When $U<U_{c1}$, the system is metallic with a finite density of states 
around the Fermi level. 
If the system enters the superfluid state,
weakly coupled Cooper pairs are formed, which give rise to
a tiny superfluid gap together with coherence peaks at its edges.
This characteristic behavior
is found for all the band fillings considered,
so we can say that a BCS-type superfluid state is realized 
in the region $(U\sim 1)$.
By examining the critical behavior of the pair potential,
we deduce $U_{c1}\sim 0.62, 0.64$ and $0.68$ for 
$n=0.5, 0.25$ and $0.125$, respectively.

When $U\sim 5$, the pair potential 
has a maximum and the double occupancy approaches
the particle density $(d\sim n)$, as shown in Fig.~\ref{fig:50}.
This means that most fermions are tightly coupled 
to form paired bosons, which stabilizes the superfluid ground state.
In this case, the energy scale should be $t^2/U$, characteristic of 
the hopping for the paired bosons.
Therefore, a further increase in the attractive interaction 
effectively increases the temperature of the system,
making the superfluid state unstable.
Eventually, the pair potential vanishes and
a phase transition occurs to the normal state 
at another critical point $U_{c2}$,
as shown in Fig.~\ref{fig:50}~(a).
In contrast to the BCS region,
a large gap remains in the density of states
and the phase transition little affects its features,
as shown in Fig.~\ref{fig:dos-50}.
This originates from the fact that in the region $(U\gtrsim 5)$, 
paired bosons exist 
in the normal state as well as in the superfluid state.
Therefore, we conclude that a BEC-type superfluid state,
which can be regarded as the condensation of paired bosons, 
is stabilized at low temperatures.
The critical values are obtained as 
$U_{c2} \sim 12.5, 11.4$ and $9.7$ for $n=0.5, 0.25$ and $0.125$.

\begin{figure}[htb]
\begin{center}
\includegraphics[width=8.5cm]{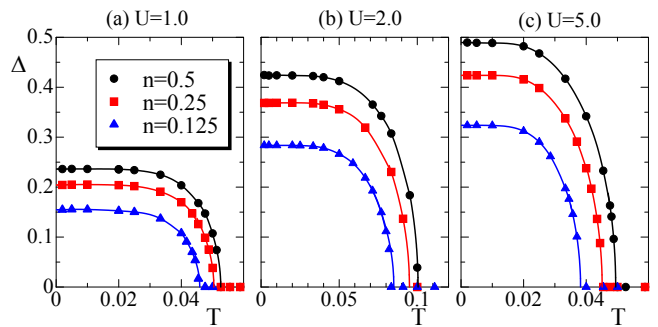}
\caption{Pair potential as a function of temperature in systems with 
$U=1.0$ (a), $2.0$ (b), and $5.0$ (c).
Circles, squares, and triangles represent the results for 
the systems with $n=0.5$, $0.25$ and $0.125$, respectively.
}
\label{fig:Delta125}
\end{center}
\end{figure}

\begin{figure}[htb]
\begin{center}
\includegraphics[width=7cm]{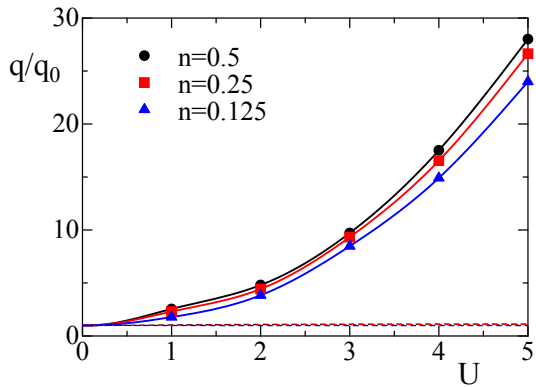}
\caption{The ratio $q/q_0$ as a function of the interaction $U$. $q_0$ 
is the universal value in the weak-coupling limit obtained from 
the BCS theory.
Circles, squares, and triangles represent the results for 
the system with $n=0.5$, $0.25$ and $0.125$, respectively.
}
\label{fig:UDTc}
\end{center}
\end{figure}

\begin{figure}[b]
\begin{center}
\includegraphics[width=7cm]{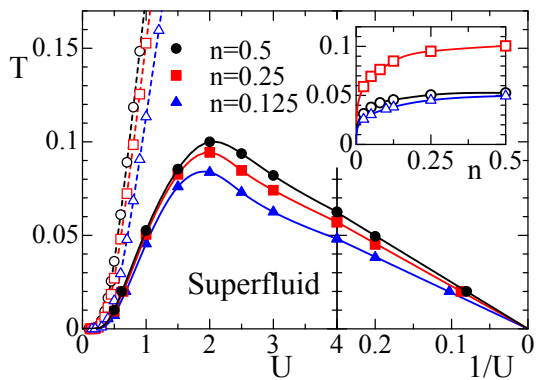}
\caption{Phase diagrams for the attractive Hubbard model.
Circles, squares, and triangles indicate the phase boundaries 
in the systems with $n=0.5, 0.25$ and $0.125$.
Solid (open) symbols are the results obtained by DMFT with CTQMC 
(the BCS theory). Lines are guides to eyes.
Inset shows that the critical temperature as a function of the particle density.
Open circles, squares, and triangles are the results for $U=1, 2$ and $5$.
}
\label{fig:PD}
\end{center}
\end{figure}

In the superfluid phase, the BCS-type state $(U\sim 1)$
is adiabatically connected to the BEC-type one $(U\gtrsim 5)$.
A BCS-BEC crossover thus occurs between the two states. 
Figure~\ref{fig:Delta125} shows the temperature dependence of 
the pair potential for three cases with $U=1, 2$ and $5$.
When the temperature is decreased,
the pair potential is induced at the critical temperature,
where the phase transition occurs from the normal state to the superfluid state.
We find that  
the pair potential increases monotonically with decreasing temperature and 
saturates at low temperatures $(T\lesssim 0.01)$.
Extrapolating the pair potential to zero temperature $\Delta_{T=0}$,
we also deduce the quantity $q =U\Delta_{T=0}/T_c $, 
as shown in Fig.~\ref{fig:UDTc}.
In the weak coupling limit ($U\rightarrow 0$), 
a BCS-type superfluid state is realized, where
this quantity is independent of the band filling and 
takes the universal value $q_0=1.764$,
according to the simple BCS mean-field theory.
Increasing the interaction strength, $q$ monotonically increases,
in a way which depends on the band filling.
In the strong coupling region, 
the quantity is proportional to the square of the interaction strength
$q \sim U^2$ since the pair potential $\Delta_{T=0}$ should saturate
and $T_c \sim 1/U$.
We note that these results differ considerably from those obtained by 
the simple BCS mean-field theory, where
quantum fluctuations are not taken into account properly. 
In fact, the critical temperature is always overestimated 
(as shown in Fig. \ref{fig:PD}) and $q_{BCS}$ is almost constant, 
as shown by the dashed lines in Fig.~\ref{fig:UDTc}.
Therefore, 
it is crucial to take into account dynamical correlations correctly 
in the attractive Hubbard model with $U\gtrsim 1$.

By performing similar calculations, we determined the phase diagram
 shown in Fig.~\ref{fig:PD}.
In the weak coupling region, the BCS-type superfluid state is realized. 
On the other hand, the BEC-type superfluid state is realized 
in the strong coupling region, where the phase boundary scales 
with the inverse of the interaction, as discussed above.
The critical temperature reaches 
its maximum value in all cases $n=0.5, 0.25$ and $0.125$
in the intermediate region $(U\sim 2)$,
where the BCS-BEC crossover occurs.
We also find that a decrease of the particle density
monotonically shifts the phase boundary.
The inset of Fig.~\ref{fig:PD} shows the particle density 
dependence of the critical temperature in systems with 
$U=1, 2$ and $5$. 
It is found that the critical temperature 
is slightly decreased away from half filling.
When the particle density approaches the dilute limit $(n\rightarrow 0)$, 
the critical temperature rapidly decreases, as
$T_c \sim n^\alpha$, where $\alpha= 0.2\sim 0.3$.

In this section,
we have discussed how the superfluid state is realized 
at low temperatures and determined the phase diagram of 
the attractive Hubbard model. 
In the following section, we focus on the dilute system 
to examine dynamical properties in the BCS-BEC crossover region.
We then clarify how the gap structure appears in the density of states.

\section{Low-temperature properties near the BCS-BEC crossover}

In the section, we study low temperature properties 
in the intermediate coupling region, where the BCS-BEC crossover occurs.
It is known that in the dense system,
the BCS-BEC crossover occurs at an interaction which is 
slightly smaller than the value corresponding to the maximum $T_c$ \cite{Toschi} 
and no pseudogap appears near the phase boundary~\cite{KogaQMC1}.
On the other hand, 
it has been reported that a pseudogap always appears 
at the critical temperature
in the Fermi gas system~\cite{Watanabe}.
Therefore, it is necessary to clarify how the pseudogap behavior is realized  
in the dilute system.
To clarify this, we focus on the attractive Hubbard model 
with a low particle density ($n=0.125$) and 
compute spectral functions in the BCS-BEC crossover region.

\begin{figure}[htb]
\begin{center}
\includegraphics[width=7cm]{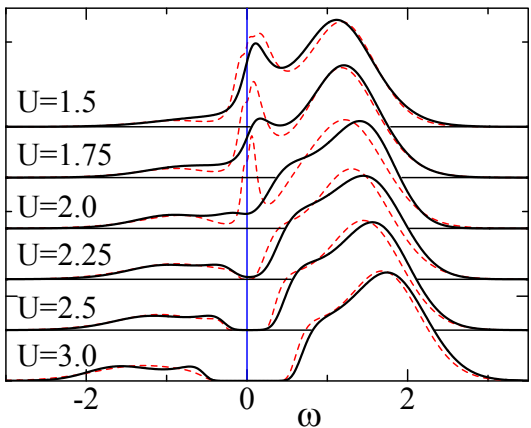}
\caption{Solid lines indicate the density of states at $T=0.087$ 
in the system with $n=0.125$.
Dashed lines indicate the results at $T=0.02$ in a system  
which is restricted to be paramagnetic.
}
\label{fig:rho12}
\end{center}
\end{figure}

We note that the gap structure appears even in the normal state 
if the preformed pairs are stabilized by the large attractive interaction.
This means that it is not directly related to the realization of 
the superfluid state, but rather to the crossover 
between the metallic and insulating state, which occurs in the normal phase. 
To make this clear,
we first examine the interaction dependence of dynamical properties 
at the high temperature $T=0.087$.
Figure~\ref{fig:rho12} shows the density of states 
for the system with $n=0.125$.
We clearly find double peaks in the case $U=1.5$, where
the metallic state is realized. 
A remarkable point is that a large density of states appears 
around the Fermi level.
This is reminiscent of heavy fermion behavior in the repulsive Hubbard model,
where a quasi-particle peak develops in the metallic state
close to the Mott transition. 
In the attractive case, 
if the ground state is restricted to be paramagnetic,
a pairing transition occurs between the metallic and insulating states 
at any filling~\cite{Capone}.
The results in the paramagnetic state at a lower temperature $(T=0.02)$ 
are shown as the dashed lines in Fig.~\ref{fig:rho12}.
It is found that the sharp quasi-particle peak grows in the region $U\lesssim 2$.
Therefore, we conclude that the large density of states near the Fermi level,
which appears above the critical temperature at $U=1.5$, 
results from particle correlations. 
We wish to note that this behavior is characteristic of the lattice model,
in contrast to the interacting Fermi gas system where a
gap behavior appears at the critical temperature~\cite{Watanabe}.

Further increase in the attractive interaction leads to the crossover 
from the heavy metallic state to the insulating state.
It is found that the quasi-particle peak collapses and 
a dip structure appears instead around the maximum $T_c$ value $(U\sim 2.0)$.
Its width continuously grows with increasing interaction, 
as shown in Fig.~\ref{fig:rho12}.
Thus, strong pairing correlations stabilize
the gap structure in the vicinity of the Fermi level.

In the intermediate coupling region 
the crossover between the metallic and insulating states occurs, 
which is associated with the pairing transition in the normal state, 
as discussed above.
It is known that at zero temperature,
this transition point is shifted away from half filling, {\it e.g.}
$U_c\sim 3.0, 2.4$ and 1.12 
in the cases with $n=0.5, 0.25$ and $0_+$~\cite{Capone}.
Therefore, we can say that the low energy properties 
around the critical temperature
depend on the particle density as well as the interaction strength.
This should have important implications for the observation of the pseudogap 
behavior in fermionic optical lattices, where 
these parameters can be controlled experimentally.
Namely, in the BCS-BEC crossover region,
heavy fermion behavior appears in the dense system~\cite{KogaQMC1}, while
a dip structure (pseudogap behavior) is found in the dilute system.

\begin{figure}[htb]
\begin{center}
\includegraphics[width=8cm]{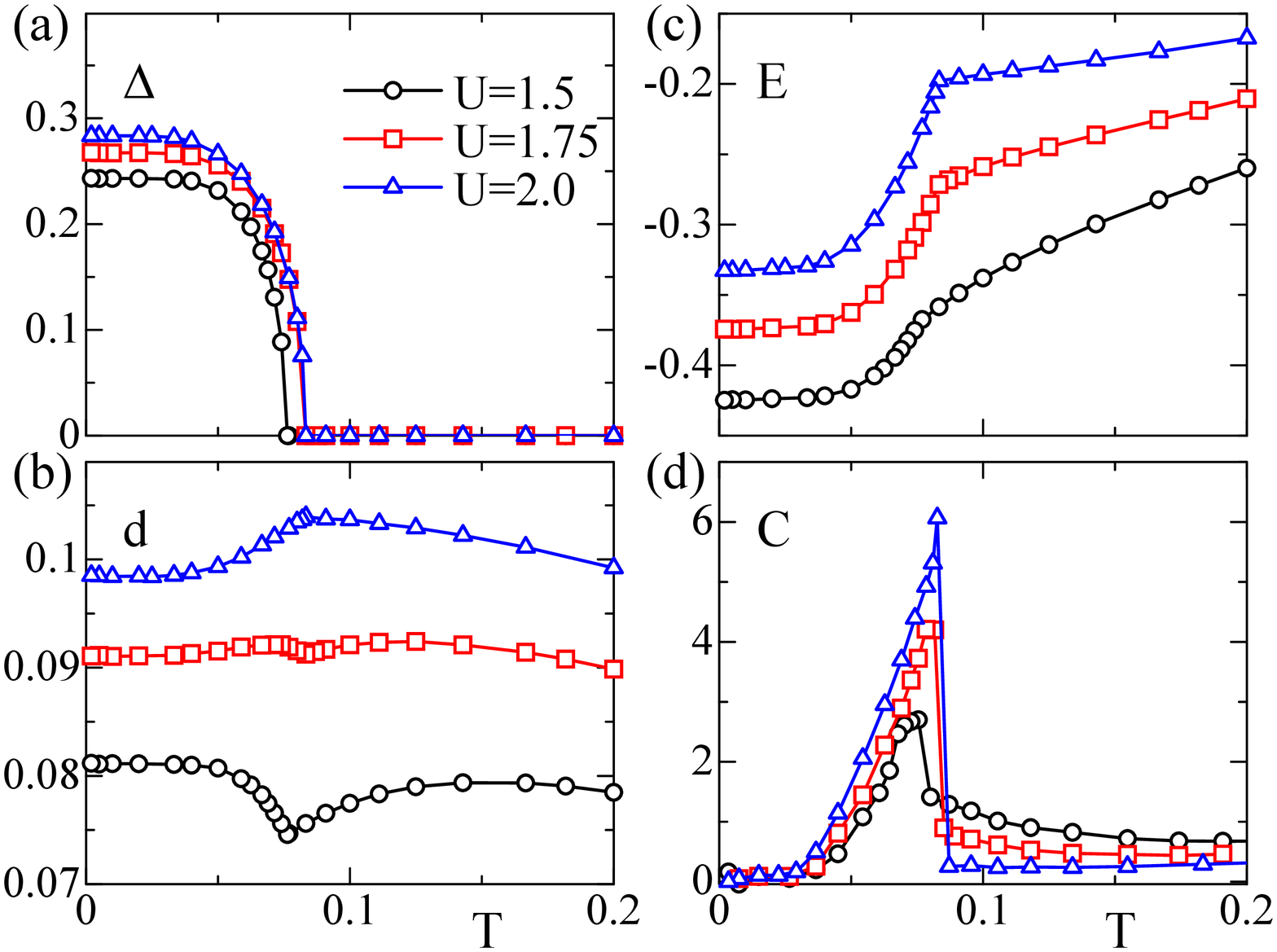}
\caption{The pair potential (a), the double occupancy (b),
the internal energy (c), and the specific heat (d) 
as a function of temperature around the BCS-BEC crossover.
}
\label{fig:E}
\end{center}
\end{figure}

\begin{figure}[b]
\begin{center}
\includegraphics[width=8cm]{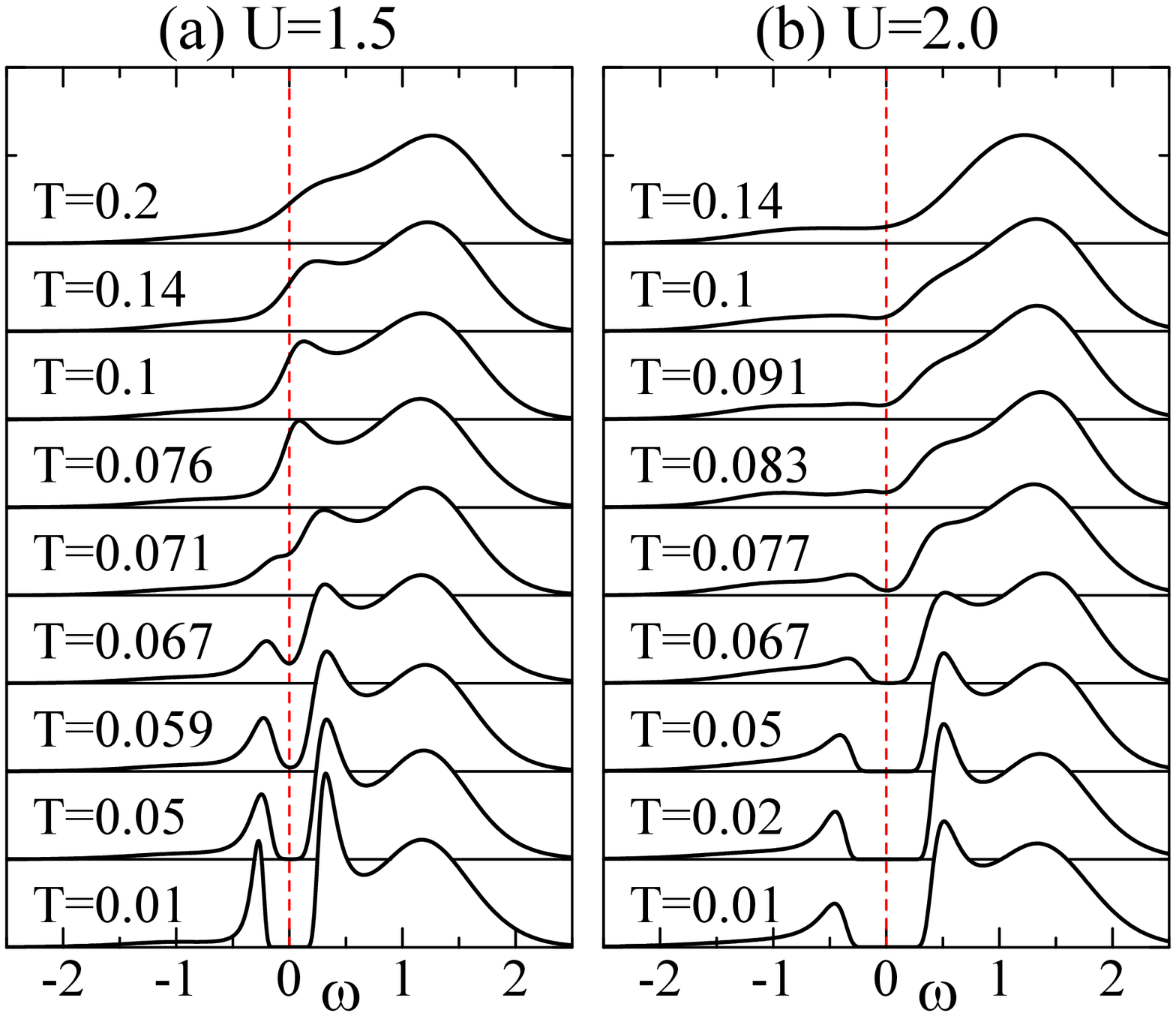}
\caption{Density of states in the dilute system ($n=0.125$)
with $U=1.5$ (a) and $2.0$ (b), where $T_c \sim 0.076$ and $0.083$, respectively.
}
\label{fig:dos-2}
\end{center}
\end{figure}

Let us consider the temperature dependence 
in the dilute system in the BCS-BEC crossover region.
We calculated the pair potential, double occupancy,
internal energy, and the specific heat, as shown in Fig.~\ref{fig:E}. 
At the critical temperature, the pair potential is induced, 
and a cusp singularity appears in the curves of 
the double occupancy and the internal energy.
A jump singularity appears in the curve of the specific heat,
and its height increases with increase of the interaction.
We note that the cusp singularity in the double occupancy vanishes
at a certain interaction $U^*(\sim 1.75)$, {\it i.e.} 
$\Delta d/\Delta T|_{T=T_C+\delta}=\Delta d/\Delta T|_{T=T_C-\delta}$.
This is due to the crossover between the heavy metallic state 
and the insulating state.
When the temperature is decreased in the case $U<U^*$, 
the double occupancy reaches a maximum at a certain temperature 
$T_{max}$ and is decreased as temperature is lowered to the critical temperature.
This suggests the formation of the heavy fermion state 
in the region with $T_c < T < T_{max}$.
In fact, we find 
an enhancement of the quasi-particle peak
in Fig. \ref{fig:dos-2} (a).
On the other hand, when $U>U^*$, 
the attractive interaction stabilizes preformed pairs at high temperatures
and thereby the decrease in the temperature 
simply increases the double occupancy.
Thus the dip structure in the density of states is stabilized
above the critical temperature, as shown in Fig.~\ref{fig:dos-2}~(b).

Below the critical temperature, the static quantities change gradually,
as shown in Fig. \ref{fig:E}.
When $U=1.5$, the decrease of the temperature increases the pair potential,
where the heavy quasi-particle peak collapses and 
the superfluid gap opens around the Fermi level, 
as shown in Fig. \ref{fig:dos-2} (a).
On the other hand, when $U=2.0$,
the dip structure found at the critical temperature
smoothly evolves into the gap structure at low temperatures.
This behavior is also observed in the system with 
a lower particle density $n=0.025$
at the temperature $T/T_c>0.17$. 
Therefore, we can say that the region with a dip structure at high temperatures
is adiabatically connected to that with the superfluid gap 
at low temperatures.
This is in contrast to the results 
for the three-dimensional Fermi gas system~\cite{Watanabe}.
In the gas system, decreasing the temperature around the BCS-BEC crossover 
smears out the pseudogap structure, and the superfluid gap develops 
below a certain temperature.
This may suggest that
$k$-dependent correlations plays a crucial role 
in realizing these dynamical properties. 
This topic is beyond the scope of our paper,
but it is important to clarify how two kinds of gap structures appear  
in a low dimensional optical lattice system,
by taking into account spatial correlations as well as dynamical correlations.

\section{Summary}
We have investigated the attractive Hubbard model in infinite spatial dimensions 
by combining dynamical mean-field theory with 
a strong-coupling continuous-time quantum Monte Carlo method.
By calculating the superfluid order parameter and the double occupancy, 
we have systematically studied the stability of 
the superfluid state and determined the phase diagram of the system.
By computing the density of states,
we have found that 
the gap structure is strongly affected 
by the interaction strength and the particle density,
which is associated with the pairing transition in the normal state.
Namely, around the BCS-BEC crossover, 
a dip structure appears in the dilute system 
while heavy fermion behavior is found in the dense system.
We have also examined the dynamical properties in the superfluid state
and have clarified that the dip structure above the critical temperature
continuously evolves into the superfluid gap 
with decreasing temperature.

In this paper, we have considered the dynamical properties 
characteristic of the infinite-dimensional lattice model,
which are qualitatively different from those in the Fermi gas system.
It is an interesting problem 
to clarify how the low energy properties are changed
by adding the lattice potential to the Fermi gas system,
which is now under consideration.

\section*{Acknowledgments}
The authors thank Y. Ohashi and Th. Pruschke for valuable discussions. 
This work was partly supported by the Grant-in-Aid for Scientific Research 
20740194 (A.K.) and 
the Global COE Program ``Nanoscience and Quantum Physics" from 
the Ministry of Education, Culture, Sports, Science and Technology (MEXT) 
of Japan. PW acknowledges support from SNF Grant PP002-118866.
The simulations have been performed using some of 
the ALPS libraries~\cite{alps1.3}.

\section*{Appendix}

In this study, we have used the strong-coupling version of the CTQMC method 
to solve the effective Anderson impurity model \cite{CTQMC}. 
This method is based 
on a stochastic sampling of an expansion of the partition function in powers
of the impurity-bath hybridization. 
A given Monte Carlo configuration can be represented 
by a collection of segments $\{\tau_{i\sigma}', \tau_{i\sigma}\}$, where
$\tau_{i\sigma}' (\tau_{i\sigma})$ is the starting (end) time for
the $i$-th segment with spin $\sigma$. These segments represent time intervals
in which an electron of spin $\sigma$ resides on the impurity. The Monte Carlo 
simulation proceeds via local updates, such as insertion/removal of a segment 
or empty space between segments (called anti-segment),
or shifts of segment end-points.  

Here, we discuss updates which improve the sampling efficiency 
in the strong-coupling region.
For $|U|\gtrsim 3$, the large energy cost for inserting or removing 
(anti-) segments leads to a
high rejection rate for proposed insertion/removal updates. 
In fact, the local (impurity) contribution to the Monte Carlo weight is given by 
\begin{eqnarray}
w_{loc}=\exp\left[-E_f\sum_{\sigma} l_{\sigma} +Ul_{overlap}\right],
\end{eqnarray}
where $E_f$ is the energy level at the impurity site
and $l_{\sigma}$ and $l_{overlap}$ are the total lengths of 
$\sigma$ segments, and the overlap between up and down segments.
Thus, in the strong-coupling regime, at low temperature, 
the acceptance probability
for a generic update will be exponentially suppressed.
One possibility is to take this exponential dependence into account 
on the level of the proposal probabilities. 
Another possibility to overcome this bottleneck is to consider updates
which change the configuration for both spins simultaneously.
When both spins are flipped between occupied and unoccupied states
in a certain time interval,
the energy change on the impurity site is not so large. 
In particular, the corresponding
energy change is zero at half filling $(E_f = U/2)$.

\begin{figure}[htb]
\begin{center}
\includegraphics[width=7cm]{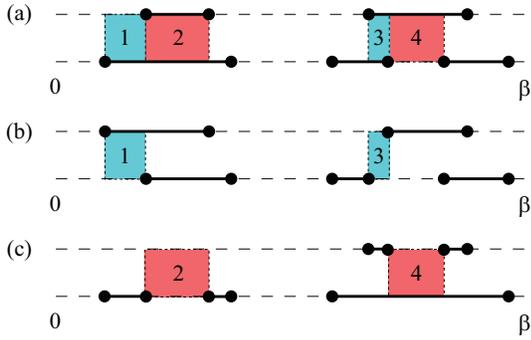}
\caption{Efficient updates in the strong-coupling regime. 
The two imaginary time intervals correspond to spin up and down.
Bold lines represent segments, {\it i.e.} a time interval,
in which an electron resides on the impurity.
(a) The shaded regions indicate examples of time intervals for which 
an improved updated will be proposed.
(b) Resulting configuration if the update for time intervals 1 and 3 
is accepted. 
These updates correspond to two shifts of segment end-points. 
(c) Resulting configuration if the update for intervals 2 and 4 is accepted. 
Here, the update corresponds to two insertions/removals of a segment/antisegment.
}
\label{fig:update}
\end{center}
\end{figure}
We discuss here explicitly one of the simplest of these updates,
which is schmatically shown in Fig.~\ref{fig:update}.
We first choose a random time interval, defined by a pair of 
neighboring creation/annihilation operators 
$\{\tau_{i\sigma}, \tau_{i\sigma}'\}$.
Note that the spin and type (creator/annihilator) is arbitrary, so the 
time interval need not correspond to any segment.
The proposal probability is given by 
$p^{prop}=(2N)^{-1}$
where $N$ is the total number of segments.
The update consists of flipping both spins in the time interval.
Note that this process is composed of two standard updates, {\it i.e.}
two insertions/removals of a segment/antisegment, or
two shift updates. 
Therefore, it is easy to add these updates to an existing CTQMC code.
The above updates were found to improve the efficiency of 
Monte Carlo simulations in the strong coupling regime.

\end{document}